\documentclass[aps,prl,twocolumn,showpacs]{revtex4}
\usepackage{amssymb}
\usepackage{txfonts}
\usepackage{graphicx}

\begin{document}

  %draft

\title{Linear Temperature Dependence of the Lower Critical Field $H_{c1}$ in F-Doped LaOFeAs Superconductors}

\author{Cong Ren$^*$, Zhao-Sheng Wang, Huan Yang, Xiyu Zhu, Lei Fang, Gang Mu, Lei Shan, and Hai-Hu Wen}
\affiliation{National Laboratory for Superconductivity, Institute of
Physics and Beijing National Laboratory for Condensed Matter
Physics, Chinese Academy of Sciences, P.O. Box 603, Beijing 100190,
China}

\begin{abstract}
We present the first experimental results of the lower critical
field $H_{c1}$ of the newly discovered F-doped superconductor
LaO$_{0.9}$F$_{0.1}$FeAs (F-LaOFeAs) by global and local
magnetization measurements. It is found that $H_{c1}$ showed an
clear linear-$T$ dependence down to a temperature of 2 K, indicative
of an unconventional pairing symmetry with a nodal gap function.
Based on the $d$-wave model, we estimated a maximum gap value
$\Delta_0=4.0 \pm 0.6$ meV, in consistent with the recent specific
heat and point-contact tunneling measurements. Taking the
demagnetization factor into account, the absolute value of
$H_{c1}(0)$ is determined to be about 54 Oe, manifesting a low
superfluid density for LaO$_{0.9}$F$_{0.1}$FeAs.
\end{abstract}


 \maketitle

The recent discovery of novel superconductivity in rare-earth
iron-based layered superconductors has received great attention in
the scientific community
\cite{Jap,ChenGF,Zhu,Oak,Wen1,Chen,Wang,Ren2,Ren,Peng,Gurevich,Dai,Du,Mazin,Fang,Cheny}.
Except for the cuprate superconductors, these new materials
ReO$_{1-x}$F$_x$FeAs exhibit quite high critical temperatures:
26-28K for Re=La at $x$=0.05-0.11\cite{Jap,ChenGF,Zhu,Oak}, 36K for
Re=Gd at $x$=0.17\cite{Peng}, 41K for Re=Ce with
$x$=0.16\cite{Wang}, 43K for Re=Sm with $x$=0.15\cite{Chen}, and 52K
for Re=Nd, Pr at $x$=0.15\cite{Ren2,Ren}, as well as 25K in hole
doped La$_{1-x}$Sr$_x$OFeAs \cite{Wen1}. One of the crucial issues
to understand the underlying superconducting mechanism in such
transition metal-based systems is the pairing interaction,
\textit{i.e.}, the symmetry of the superconducting order parameter
and the nature of the low energy excitations. Up to now, there have
been several investigations on the pairing symmetry for
LaO$_{0.9}$F$_{0.1}$FeAs (F-LaOFeAs) superconductor. Recent specific
heat measurement has revealed a nonlinear magnetic field dependence
of the electronic specific heat coefficient $\gamma(H)$ in the low
temperature limit, which is consistent with the prediction for a
nodal superconductor \cite{Wen2}. The presence of a node in the
superconducting gap has also been verified by the observation of a
zero-biased conductance peak in point contact tunneling spectroscopy
\cite{Shan}.  On the other hand, an extended $s$-wave multi-band
superconductivity was proposed \cite{Gurevich} to account for the
experiment results of the extremely high upper critical field in
F-LaOFeAs. From the band structure point of view, the two-band model
has been strengthened with $d$-wave pairing interactions originated
from the intra-band antiferromagnetic coupling plus an effective
inter-band antiferromagnetic interaction \cite{Cheny}. Because the
current experiments have been performed on polycrystalline samples,
the experimental results within the context of the symmetry of the
pair state is not yet well determined.

Lower critical field $H_{c1}(T)$, or magnetic penetration depth
$\lambda(T)$ are fundamental probes of the existence of nodes, or
two-gap in the superconducting gap function of unconventional
superconductors.  As an advantage, $H_{c1}(T)$ measurement probe
relatively large distances ($\lambda \sim 1000$ \AA) and are far
less sensitive to sample surface quality.  In this work we present
the first detailed measurements of the temperature dependence of the
lower critical field $H_{c1}$ of superconducting
LaO$_{0.9}$F$_{0.1}$FeAs polycrystals. We found a predominately
linear-$T$ dependence of $H_{c1}$ down to a temperature of 2 K,
which is a strong evidence for nodes in the superconducting gap of
F-LaOFeAs samples.

\begin{figure}
\includegraphics[scale=0.75]{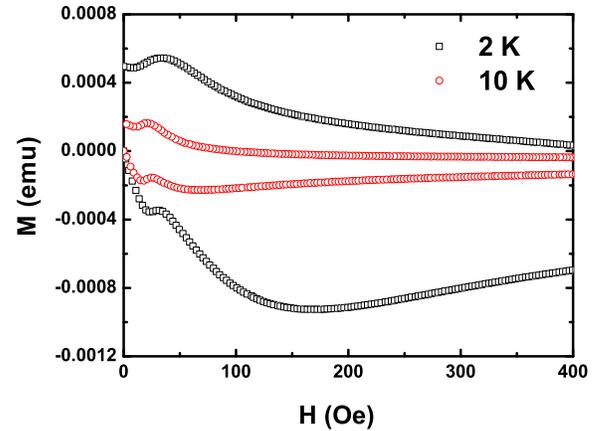}
\caption{\label{fig:fig1}(Color online) Magnetization hysteresis
loops of LaO$_{0.9}$F$_{0.1}$FeAs measured by VSM in ZFC mode at
$T=2, 10 $ K.}
\end{figure}

Our polycrystalline samples were synthesized by using a two-step
solid state reaction method. The superconducting transition
temperature $T_c$ defined as the onset of the drop in resistivity
was 26 K with a transition width of $\Delta T_c=3$ K (10\%-90\% of
normal state resistivity). The details of the sample synthesis and
the characterization have been reported in our previous papers
\cite{Wen2,Zhu}.  Our samples were relatively dense with a metallic
shiny grain surface and shaped into various dimensions. The data
presented here were taken on a platelike polycrystal with the
dimension of $0.6\times 0.6 \times 0.2$ mm$^3$.

Global dc magnetization measurement were carried out by a high
resolution vibrating sample magnetometer (VSM) (Quantum Design) with
applied field $H$ parallel to the shortest lateral side of the
sample. Local magnetization loops were measured using a two
dimensional electron gas (2DEG) micro Hall probe with an active area
of $10\times 10\ \mu$m$^2$.  The Hall probe was characterized
without sample attachment at different temperatures. In our
experiment, all $M(H)$ curves were taken in zero field cool (ZFC)
mode with initial temperature up to 40 K. To minimize the complex
effects of the character of the field penetration in layered
structure, such as Bean-Livingston surface barriers and/or
geometrical barriers \cite{Zeldov,Kes}, we used a low field sweep
rate of 30 Oe/min to measure isothermal magnetization curves.

Typical global magnetization loops $M(H)$ of our specimen at $T$=2
and 10 K are presented in Fig. 1. It is seen that the $M(H)$ curves
are symmetric and show a relatively large critical current density
in the mixed state using Bean critical state model.  This implies
that the surface barriers for flux entry does not play an important
role in our case since Bean-Livingston surface barriers are expected
to give rise to a very asymmetric magnetization loop \cite{K}, and
the pinning mechanism mainly arises from bulk pinning. In addition,
a feature of a peak in $M(H)$ appears followed the full penetration
field. This so called ``second peak'' in $M(H)$ loops has been
widely observed in strongly layered superconductors, such as
Bi$_2$Sr$_2$CaCu$_2$O$_8$ and NdCeCuO single crystals, and
attributed to the vortex phase transition to vortex glass state
\cite{Zeldov1,Zeldov2}. Here the observation of second peak effect
in our specimen indicates that the LaO$_{0.9}$F$_{0.1}$FeAs is
strongly anisotropic and the intergrain links are strong,
representative of the bulk property.

\begin{figure}
\includegraphics[scale=0.9]{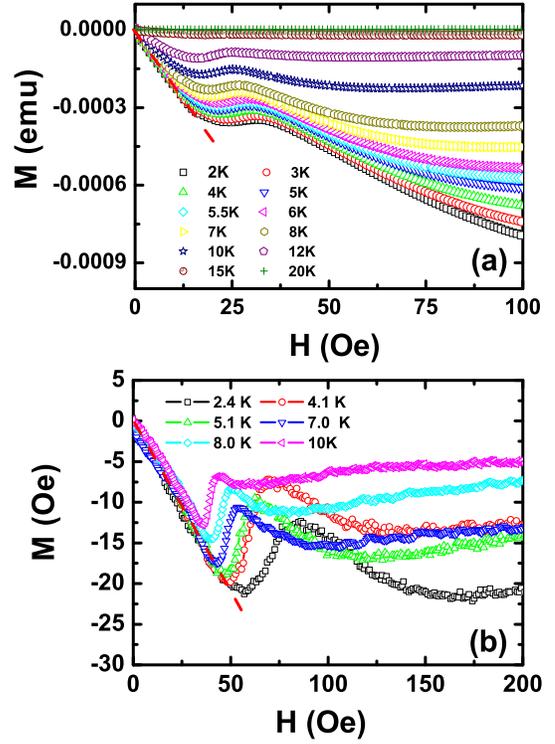}
\caption{\label{fig:fig2}(Color online) The magnetization curve
$M(H)$ by VSM (a) and Hall probe (b), at various temperatures,
respectively. The dotted red lines are the ``Meissner line'' showing
the linearity of these curves at low fields. }
\end{figure}

Figure 2(a) and Figure 2(b) are the initial isothermal $M(H)$ curves
measured by VSM and Hall probe over the temperature range from 2 to
20 K, respectively. The second peak effect is more distinct in the
curves by local measurement, as shown in Fig. 2(b).  These $M(H)$
curves show clearly a linear dependence of the magnetization on
field caused by Meissner effect at low fields. In global
measurement, due to the bulk pinning the entry of vortices is
gradual, comparing to the local measurement. This causes a gradual
$M(H)$ curve, as shown in Fig. 2(a).  It is difficult to determine
$H_{c1}$ using these $M(H)$ curves. Therefore we use the method in
Ref. 24 to extract the value of $H_{c1}$: According to the Bean
critical state model for type-II superconductors, magnetic induction
$B$ induced by the external field is:
\begin{equation}
B=A(H-H_{c1})^2/H^*  \ \ \ \ \ \ (H_{c1} < H < H^*),
\end{equation}
where $A$ is a sample-dependent constant and $H^*$ is related to
critical current density.  According to this model, a plot of
$B^{1/2}$ vs $H$ should yield a straight line with a threshold at
$H_{c1}$. Fig. 3 shows such plot of $B^{1/2}$ vs $H$ at $T=2, 3, 4 $
and 10 K for the global curves, where $B^{1/2}=(M-M_{ML})^{1/2}$ and
$M_{ML}$ is the Meissner line for each curve, shown as the dotted
line in Fig. 2 for the curve of $T=2$ K.  As displayed, the data are
well described by Eq. (1) except a small kink at $H > H_{c1}$.  This
kink is due to the second peak effect in the original $M(H)$ curves.
Neglecting this effect, $H_{c1}$ is well determined as the field of
the threshold of $B$, as demonstrated in Fig. 3.  By the same
procedure, we also determined the $H_{c1}(T)$ data from the $M(H)$
curves using Hall probe (not shown here for simplicity) .

\begin{figure}
\includegraphics[scale=0.7]{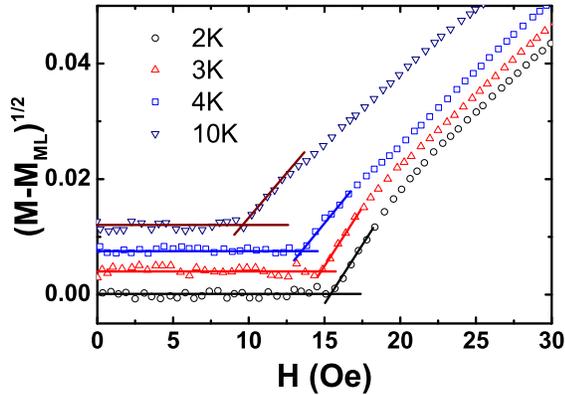}
\caption{\label{fig:fig3}(Color online) The $M(H)$ data in the form
of $(M-M_{ML})^{1/2}$ vs $H$.  $M_{ML}$ is the Meissner line shown
in Fig. 2.   The curves ($T$=3,4,10 K) are shifted up for clarity.
The colored lines are guides to the eyes. }
\end{figure}

Depicted in Fig. 4(a) are our main results, in which the extracted
$H_{c1}$ is plotted as a function of $T$ for both VSM and Hall probe
measurements.  As shown in the inset of Fig. 4(a), the $H_{c1}$
measured by VSM is much lower than those by local measurement even
though we measured the same sample.  The reason for this difference
is as follows. In global measurement, the VSM coil picks up any
signal once vortices start to enter into the specimen through the
edge of the sample. On the other hand, the local Hall probe detects
signal only when vortices almost fully penetrates into the sample
and reaches the region where the Hall probe locates.  Assuming the
Hall probe is located at the center of the sample, which is far from
the sample edge, the Hall probe detects signal only from the central
region. Accordingly, the $H_{c1}$ measured by local Hall sensor is
expected to be larger than that of global VSM measurement. It also
explains the experimental results that the $M(H)$ curves by global
measurement is more gradual around the penetration region, comparing
those measured by local magnetization measurement, which is exactly
the case we observed in Fig. 2. The fact that the demagnetization
factor is different for local and global measurements can manifest
itself by scaling the $H_{c1}(T)$ curve measured by VSM to the one
by Hall sensor. After we correct a scaling factor of 3.22 for the
$H_{c1}(T)$ by VSM, the two $H_{c1}(T)$ lines almost collapse onto
the same line except the part of high $T$, which is due to large
error bars, as shown in the main panel of Fig. 4(a). The coincidence
of the two scaled $H_{c1}(T)$ curves indicates that our results are
intrinsic in physics, independent of measuring device and technique.
Thereafter, if not specially mentioned, we choose the data obtained
by VSM for our analysis and discussions since the demagnetization
factor is well determined for global measurement, as discussed
below.

We now analyze and discuss the obtained data of $H_{c1}(T)$.
Principally one can get the absolute value of $H_{c1}(0)$ by
extrapolating the nominal $H_{c1}(T)$ curve to $T=0$, taking the
demagnetization factor into account. For samples with elliptical
cross sections, $H_{c1}$ can be deduced from the first penetration
field $H_{c1}^*$, assuming that the magnetization $M=-H_{c1}$ when
the first vortex enters into the sample.  Thus $H$ has been rescaled
to $H=H_a-NM$ and $H_{c1}=H_{c1}^*/(1-N)$, where $N$ is the
demagnetization factor and $H_a$ the external field.  For sample
with a rectangle cross section, the $M(H)$ curves can still be
approximately by a linear variation in the Meissner state. However,
the demagnetization factor has been reconsidered.  It has, for
instance, been shown by Brandt that a bar with a rectangle cross
section, the first penetration field $H_{c1}^*$ and $H_{c1}$ have
the relation \cite{Brandt}: $H_{c1}= H_{c1}^*/ \tanh\big(\sqrt{0.36
b/a}\big)$, where $a$ and $b$ are the width and the thickness of the
samples, respectively. Using this formula, we estimate the effective
demagnetization factor $N_{eff}\simeq 0.67$ as we take $a\simeq 0.6$
mm and $b\simeq 0.2$ mm for our sample. This estimated $N_{eff}$ is
consistent with those of MgB$_2$ platelets with the similar size
ratio \cite{Lyard,Shi}, in which $N_{eff}\simeq 0.6-0.69$ has been
determined.  $H_{c1}^*(0)=18$ Oe is obtained by linear fitting to
the data of $H_{c1}(T)$ below 5 K. With $N_{eff}$, the formula
$H_{c1}(0)=H_{c1}^*(0)/(1-N_{eff})$ yields a true $H_{c1}(0)=54$ Oe.
A striking feature of the nominal $H_{c1}(T)$ shows an approximate
linear-$T$ dependence from around 6 K down to 2 K, the lowest
temperature we can reach in this experiment, as shown by the
straight line in the inset of Fig. 4(a).  This approximate
linear-$T$ behavior is also evidenced from the $H_{c1}(T)$ curve
using Hall sensor.

In order to investigate the properties of superconducting gap using
the obtained linear-$T$ dependence of $H_{c1}$ in the low-$T$
region, we use the formula to calculate the absolute value of the
penetration depth $\lambda$ (taking the demagnetization effect into
account):
\begin{equation}
H_{c1}=\frac{\Phi_0}{4\pi \lambda^2}\big(\ln \kappa + 0.5 \big),
\end{equation}
where $\Phi_0=he/2e=2.07\times 10^{-7}$ Oe cm$^2$ is the flux
quantum, and $\kappa$ is the Ginzberg-Laudau parameter. Assuming
$\kappa\sim 100$, we deduce a value of $\lambda\simeq 390$ nm with
the calculated parameter $H_{c1}(0)=54$ Oe.

\begin{figure}
\includegraphics[scale=0.9]{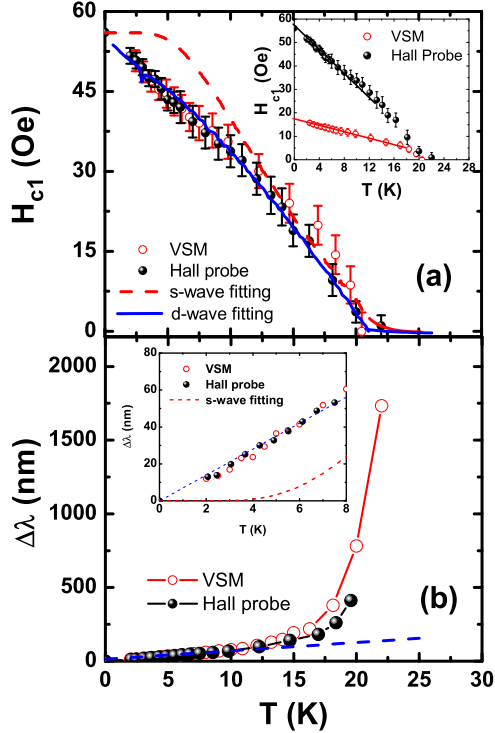}
\caption{\label{fig:fig4}(Color online) (a) $T$ dependence of
$H_{c1}$ determined from the points deviating from the linearity of
the initial $M(H)$ curves by VSM and Hall Probe measurements. The
data of $H_{c1}(T)$ by VSM is scaled by multiplying a number of 3.22
to account for the different demagnetization effects.  The solid
blue line is a $d$-wave weak-coupling BCS fit with $\Delta_0=3.85$
meV; the dashed red line is an $s$-wave weak-coupling fit. Inset:
the original data of $H_{c1}(T)$ measured by VSM and Hall Probe,
respectively. The solid lines are the linear fit to the data of low
$T$. (b) $T$ dependence of $\Delta\lambda$ calculated with the data
of $H_{c1}(T)$ and Eq. (2). Inset: the enlarged low temperature
part, which shows a clear linear-$T$ dependence of $\Delta\lambda$.
The dotted blue line is the linear fit, and the dashed red line is
an $s$-wave exponential-$T$ line. }
\end{figure}

In Fig. 4(b) we plot the deviation $\Delta
\lambda=\lambda(T)-\lambda(0)$ as a function of $T$ using
$\lambda(0)=390$ nm.  At low temperatures $T < 5$ K, $\Delta
\lambda$ shows a linear function of $T$, i. e. $\Delta\lambda\propto
T$, which is more explicit in the inset of Fig. 4(b).  This linear
dependence of $\Delta\lambda$ contradicts the result of an isotropic
$s$-wave gap superconductivity.  For an isotropic $s$-wave
superconductor, such as conventional superconductor, the low
excitation of the finite energy gap is a type of an exponentially
thermal activation.  In the isotropic $s$-wave BCS theory,
$\Delta\lambda = \lambda(0) \sqrt{\pi \Delta_0/2T}
\exp\big(-\Delta_0/T\big)$, where $\Delta_0$ is the maximum gap
value.  This theory can not describe our $\Delta\lambda(T)$ since
the exponential term gives $d\lambda(T)/dT\simeq 0$ at low $T$,
shown as the fitting line in the inset of Fig. 4(b), in contrast to
our result of $d\lambda(T)/dT \propto T$.

The linear-$T$ behavior of $\Delta\lambda$ has been observed in
high-$T_c$ superconductor YBCO \cite{Hardy,Liang},
Bi$_2$Sr$_2$CaCu$_2$O$_8$ \cite{Li,Lee} single crystals,
HgBa$_2$Ca$_2$Cu$_3$O$_{8+\delta}$ crystal powder \cite{Xiang2} and
magnetic superconductor YNi$_2$B$_2$C \cite{Oxx}. In cuprate
superconductors, such a linear-$T$ dependence of penetration depth
$\lambda$ at low $T$ is a signature for the existence of a line node
in the superconducting energy gap function, \textit{i.e.} the
$d$-wave pairing \cite{Hardy,Xiang2}, where thermal excitation of
quasiparticles near nodes in the superconducting energy gap cause
$\rho_s\equiv 1/\lambda^2$ to decline linearly with temperature.
Based on the $d$-wave pairing theory, $H_{c1}(T)/H_{c1}(0)$, i.e.,
$\lambda^2(0)/\lambda^2(T)$, is a linear-$T$ function at $T\ll T_c$,
and can be expressed as \cite{Peter,Xiang2}:
\begin{equation}
1-\frac{H_{c1}(T)}{H_{c1}(0)}=1-\frac{\lambda^2(0)}{\lambda^2(T)}=2\ln
2\frac{k_BT}{{\Delta_0}}.
\end{equation}
As shown in Fig. 4(a), the slope of the fitting line
$d(H_{c1}(T)/H_{c1}(0))/dT$ is 0.78, which yields a $\Delta_0=(1.8
\pm 0.3) k_B T_c$ based on Eq. (3).  Using the known $T_c=26$ K, we
found $\Delta_0=4.0\pm 0.6$ meV. This deduced $\Delta_0$ are in
excellent agreement with the values obtained by our recent specific
heat \cite{Wen2} and point-contact tunneling spectroscopy
\cite{Shan} measurements on similar samples.

Before conclusion, we would like to address the concern of the
effect of impurity scattering on the linear-$T$ dependence of
magnetic penetration depth. For our samples, the coherence length
$\xi_0 \sim 80$ \AA\ \cite{Wen2} and the mean free path $l>100$ \AA\
\cite{Zhu}, and thus our samples are in the moderate clean limit ($l
\sim \xi_0$). Meanwhile, we believe that nonmagnetic impurity
scattering does not strongly modify the low-$T$ properties of
F-LaOFeAs since $T_c$ of our sample, another important low-$T$
parameter, does not change much while impurity level (residual
resistivity) increases up to 2-4 times \cite{Zhu}.  This situation
also manifests itself by the fact that although different groups
adopt different methods to synthesize F-LaOFeAs superconductor with
different impurity level, the $T_c$ defined as the drop of
resistance is around 25-28 K \cite{Jap,Wen2,Wang,Oak,Gurevich}.

To summarize, we conduct both global and local magnetization
measurements on the new superconductor LaO$_{0.9}$F$_{0.1}$FeAs. The
temperature dependence of the lower critical field $H_{c1}$ is
extracted.  It is found that $H_{c1}$ show a linear-$T$ behavior at
low temperatures, suggesting a nodal gap function.  We also obtained
the lower critical field of $H_{c1}(0)=54$ Oe, showing a low
superfluid density for F-LaOFeAs. Based on the finite slope of
$H_{c1}(T)$ curve at low $T$, we estimated the maximum gap value to
be $\Delta(0)=4.0 \pm 0.6 $ meV, which is very consistent with the
results of recent specific heat and tunneling experiments on the
similar samples.

Acknowledgement: The authors are grateful to Prof. Wen-Xin Wang and
Prof. Jun-Ming Zhou for providing GaAs/AlGaAs substrates. This work
is supported by the National Science Foundation of China, the
Ministry of Science and Technology of China (973 project No:
2006CB60100, 2006CB921107, 2006CB921802), and Chinese Academy of
Sciences (Project ITSNEM).

\vspace{0.3cm}

Email address:

$^*$ cong\_ren@aphy.iphy.ac.cn

\end{document}